\documentclass[12pt]{article}
\pdfoutput=1

\usepackage{putex}
\usepackage{amsmath,amssymb,amsfonts}
\usepackage{psfrag}
\usepackage{footmisc}
\usepackage{url}
\usepackage{mathtools}
\usepackage{color}

\usepackage{tikz}
    \usepackage{amssymb,amsfonts,amsmath}
    \usepackage{tkz-euclide}
        \usetikzlibrary{arrows,calc,patterns}
\usepackage{pgfplots}

\definecolor{darkblue}{rgb}{0.1,0.1,.7}
\definecolor{purple}{rgb}{0.6,0,0.6}
\definecolor{orange}{rgb}{0.9,0.6,0}
\usepackage[colorlinks, linkcolor=darkblue, citecolor=darkblue, urlcolor=darkblue, linktocpage]{hyperref} 
\usepackage[square, comma, compress,numbers]{natbib}
\usepackage[]{amsmath}
\usepackage[]{graphicx}
\usepackage[]{latexsym}
\usepackage[utf8]{inputenc}
\usepackage{geometry}
\usepackage{amscd}
\usepackage[all,cmtip]{xy}
\usepackage{mathrsfs}

\usepackage[margin=10pt,font=small,labelfont=bf]{caption}
\usepackage{dsdshorthand}
\usepackage{changepage}
\usepackage{setspace}
\usepackage{tikz}
\usepackage{subcaption}

\usepackage[titles]{tocloft}
\usetikzlibrary{arrows,positioning}

\def\SL2{\widetilde{SL}(2,\mathbb R)}

\def\mC{\mathcal C}

\newcommand\mR{\mathbb{R}}

\numberwithin{equation}{section}

\newcommand{\grp}[1]{\mathrm{#1}}

\newcommand {\bes} {\begin {equation*}}
\newcommand {\ees} {\end {equation*}}
\newcommand {\beq} {\begin {equation}}
\newcommand {\eeq} {\end {equation}}
\newcommand {\bea} {\begin {eqnarray}}
\newcommand {\ea} {\end {eqnarray}}
\newcommand {\eea} {\end {eqnarray}}

\newcommand{\Sch}{\text{Sch}}

\numberwithin{equation}{section}

\def\<{\langle}
\def\>{\rangle}

\tikzset{
    >=stealth',
    punkt/.style={
           rectangle,
           rounded corners,
           draw=black, very thick,
           text width=15em,
           minimum height=2em,
           text centered},
    pil/.style={
           ->,
           thick,
           shorten <=2pt,
           shorten >=2pt,}
}

 \def\ie{\begin{equation}\begin{aligned}}
\def\fe{\end{aligned}\end{equation}}

\renewcommand{\=}{\; = \;}

\renewcommand{\t}{\tau}         
\newcommand{\rme}{{\rm e}}         
\newcommand{\ii}{{\rm i}}

\begin{document}

    \institution{SU}{${}^1$ Stanford Institute for Theoretical Physics, Stanford University, Stanford, CA 94305, USA}
    \institution{OX}{${}^2$ Department of Mathematics, King’s College London, The Strand, London WC2R 2LS, U.K.}
    \institution{IAS}{${}^3$  School of Natural Sciences, Institute for Advanced Study, Princeton, NJ 08540, USA}
    \institution{UW}{${}^4$ Physics Department, University of Washington, Seattle, WA, USA}

\title{
Revisiting the Logarithmic Corrections to the Black Hole Entropy
}

\authors{Luca V. Iliesiu${}^1$, Sameer Murthy$^2$ and Gustavo J. Turiaci${}^{3,4}$ }

\abstract{
Logarithmic corrections to the entropy of extremal black holes have  been successfully used to accurately match degeneracies from microscopic constructions to calculations of the gravitational path integral. In this paper, we revisit the problem of deriving such corrections for the case of extremal black holes, either non-supersymmetric or supersymmetric, and for near-extremal black holes. The zero-modes that are present at extremality are crucial, since their path integral cannot be treated quadratically and needs to be regulated. We show how the regulated result can be obtained by taking the zero-temperature limit of either the $4d$ Einstein-Maxwell or $4d$ supergravity path integral to find the Schwarzian or super-Schwarzian theories. This leads to drastically different estimates for the degeneracy of non-supersymmetric and supersymmetric extremal black holes. In a companion paper, we discuss how such zero-modes affect the calculation of BPS black holes degeneracies, using supersymmetric localization for an exact computation of the gravitational path integral.  
}
\date{}

\maketitle
\tableofcontents

\section{Introduction}
Black holes have a thermodynamic entropy that scales as one-quarter of the horizon area in Planck units in the semiclassical limit \cite{Bekenstein:1973ur, Hawking:1975vcx}. This formula receives quantum corrections, the leading effects being logarithmic in the parameters of the black hole. These logarithmic corrections to the black hole entropy can be obtained in two seemingly unrelated ways. On the one hand, such corrections can be obtained using the gravitational path integral by expanding the quantum fluctuation of all the fields in the gravitational theory to quadratic order around a given black hole background. 
On the other hand, when microscopic constructions for such black holes are available, the entropy can often  be computed exactly. Expanding the exact entropy for large black hole area gives a prediction for the logarithmic correction to the entropy.  Matching the corrections obtained in these two different ways provides an important verification that the gravitational action is indeed the low-energy effective theory for the UV complete theory used in the microscopic construction, e.g.~string theory.
In this paper we will revisit the computation of these logarithmic corrections for extremal and near-extremal black holes from the path-integral perspective, emphasizing that in the limit towards extremality corrections that are logarithmic in the temperature become, in the absence of supersymmetry, equally important to the terms logarithmic in the area. 

Upon including these corrections, the quantum entropy for extremal black holes takes the following general 
form~\cite{Banerjee:2010qc,Banerjee:2011jp,Sen:2011ba,Sen:2012cj},
\be 
    \label{eq:one-loop-result-structure-intro}
        \exp(S^\text{quant}) \= \mC S_0^{c_{\rm log}}\rme^{S_0} \left(1+O\left(\frac{1}{S_0}\right)\right)\,, \qquad S_0 \= \frac{\text{Area}}{4 G_N}\,,
\ee
where $c_{\rm log}$ is a number that is independent of the charges of the black holes and only dependent on the number of massless scalars $n_s$, spin-$1/2$ fermions $n_f$ and spin-$3/2$ fields.\footnote{As we shall explain below massive fields only contribute at subleading order in \eqref{eq:one-loop-result-structure-intro}. } In \eqref{eq:one-loop-result-structure-intro} we cannot determine the overall charge independent prefactor $\mC$ which in the path integral approach is regularization dependent and thus will be left undetermined. The area in this expression corresponds to the area of the extremal horizon.

By determining $c_{\rm log}$ in various theories of supergravity arising as the low energy limit of string theory, \cite{Banerjee:2010qc,Banerjee:2011jp,Sen:2011ba,Sen:2012cj} found agreement with the microscopic prediction. This provides a non-trivial check that $\cN=2,\,4$ or $8$ supergravity is indeed the correct low-energy description for various compactifications of string theory. Nevertheless, when supersymmetry is not preserved at extremality and the predictions for the extremal black hole degeneracy are limited,  there are reasons to question the validity of \eqref{eq:one-loop-result-structure-intro}.  Regardless of the matter content in the gravitational theory and the consequent value of  $c_{\rm log}$, \eqref{eq:one-loop-result-structure-intro} always predicts a large degeneracy for black holes at extremality. 
However, in the absence of supersymmetry in which the degeneracy of BPS states can often times be seen at weak coupling, there does not seem to be a symmetry principle that protects this degeneracy. For this reason, one might therefore expect large corrections to be present at low-temperatures to \eqref{eq:one-loop-result-structure-intro}. 

The purpose of this paper is to show how such corrections appear when integrating out the quantum fluctuations in the gravitational action around an extremal or near-extremal black hole saddle. For extremal black hole backgrounds, we
explain the treatment of zero-modes of the gravitational action which have to be integrated out exactly, in contrast to the non-zero modes in the expansion for which it suffices to work at quadratic order. 
We regulate this path integral over zero-modes by going slight away from extremality and showing that now such modes correspond to a strongly coupled mode\footnote{Though these modes are weakly coupled to all other non-zero soft-mode, they are strongly self-interacting.}---the (super)Schwarzian mode---whose path integral can luckily be performed exactly. 
For extremal non-supersymmetric black holes the integral over zero-modes vanishes, confirming that their degeneracy is much smaller than expected, {i.e.}~it does not scale as $\exp (S_0)$ but rather it is non-perturbatively smaller. For BPS  black holes we show that the integral over zero-modes is indeed finite and that the integral over all other modes produces the logarithmic correction for \eqref{eq:one-loop-result-structure-intro} previously predicted in \cite{Banerjee:2010qc,Banerjee:2011jp,Sen:2011ba,Sen:2012cj}. This thus confirms, from the path integral perspective, that the BPS black hole microstates indeed have a large degeneracy and that the logarithmic correction to this degeneracy indeed matches the string theory prediction. 
In a companion paper \cite{Iliesiu:companionPaperMicrostateCounting} we show that the integral over these zero-modes is absolutely necessary to fully reproduce microscopically predicted degeneracy when trying to localize the gravitational path integral. 

The results described above have been discussed in the past \cite{Ghosh:2019rcj, Iliesiu:2020zld,Heydeman:2020hhw,Boruch:2022tno}. The approach taken in those references was to rewrite the higher dimensional gravity theory as a two-dimensional theory in the AdS$_2$ near-horizon throat coupled to infinite fields coming from KK modes on the transverse space. The temperature dependence is then shown to be controlled by a specific mode in the metric described by (super)JT gravity \footnote{Other papers studying different aspects regarding the connection of JT gravity and near extremal black holes are \cite{Sachdev:2015efa, Almheiri:2016fws, Nayak:2018qej, Moitra:2018jqs, Castro:2018ffi, Larsen:2018cts,  Moitra:2019bub, Sachdev:2019bjn,Maldacena:2019cbz, Charles:2019tiu,Ghosh:2019rcj, David:2020ems, Larsen:2020lhg, Heydeman:2020hhw, Iliesiu:2020qvm,David:2020jhp, Castro:2021wzn, Boruch:2022tno}.}. 
The purpose of this paper is to re-derive these results using the more conventional approach of Sen and collaborators \cite{Banerjee:2010qc,Banerjee:2011jp,Sen:2011ba,Sen:2012cj} which was previously used to determine the log corrections in \eqref{eq:one-loop-result-structure-intro}. Deriving the result in this way provides a useful accounting tool for the contribution of individual modes in the gravitational path integral and corrects the approach of \cite{Banerjee:2010qc,Banerjee:2011jp,Sen:2011ba,Sen:2012cj} in a manner consistent with the JT gravity perspective advocated for in \cite{Ghosh:2019rcj, Iliesiu:2020zld,Heydeman:2020hhw,Boruch:2022tno}. 

The rest of this paper is organized as follows. In Section~\ref{sec:quantum-entropy-review} we give a more technical description of the calculation of the logarithmic correction and pinpoint the issue with the approach of~\cite{Banerjee:2010qc,Banerjee:2011jp,Sen:2011ba,Sen:2012cj} for the zero-modes that are encountered at extremality. In Section~\ref{sec:zero-modes-in-Einstein-Maxwell-theory} we discuss the distinction between zero-modes and massive modes for extremal black holes in Einstein-Maxwell theory and show how to regulate the path integral for each zero-mode that we find. Additionally, we discuss in detail the various boundary conditions that need to be imposed on the metric and gauge fields when studying black holes in the canonical and grand canonical ensembles.  In Section~\ref{sec:zero-modes-in-SUGRA} we discuss the differences between non-BPS and BPS black holes in supergravity and show that for BPS black holes the path integral over zero modes is regulated differently to give a large degeneracy at extremality. We end with a discussion about possible non-perturbative corrections to the degeneracies found at extremality.

\section{
Background and points that we address}
\label{sec:quantum-entropy-review}

The goal of this paper is to revisit in detail the derivation of the logarithmic corrections to the black hole entropy and emphasize the important role that zero-modes present at extremality play in the computation. For concreteness, throughout this paper our focus is on  extremal and near-extremal black holes in $4d$ asymptotically flat space.  
Nevertheless, our conclusions regarding the role of the zero-modes are more general and only depend on the near-horizon isometry that is preserved at extremality and thus apply regardless of the number of spacetime dimensions and the asymptotics of the spacetime. 

Our starting point is the quantum entropy as defined by Sen in \cite{Sen:2008yk,Sen:2008vm}, with a modification that we spell out below that is important for the analysis of zero-modes. 
This approach is based on the observation that near extremality the near horizon of the black holes (at least the ones we will consider) have a two dimensional AdS spacetime along the radial and time directions and an $S^2$ internal space.  In the limit of zero temperature, the quantum entropy can then be defined using the gravitational path integral with AdS$_2 \times S^2$ boundary conditions:
\be \label{qef}
\exp(S^\text{quant}) &\=
\Bigl\langle \exp\bigl(-\ii \, q_I \oint  A^I \bigr) \Bigr\rangle_{\rm{AdS}_2 \times S^2 \text{ with } T\to 0}^\text{reg} \, \nn \\ &\= \int [Dg \,D A \, D \chi\,D\Psi\dots] \,\,\rme^{-S_\text{grav, bulk} - S_\text{grav, bdy}} \,.
\ee 
This is an extension of the Gibbons-Hawking prescription \cite{Gibbons:1976ue} to the near-horizon geometry of extremal black holes including, in principle, all quantum effects. The extremal black hole whose entropy is computed through \eqref{qef} should be thought of as the zero-temperature limit of a near-extremal black hole.
We now explain how to consider this limit as well as describe all the elements that go into \eqref{qef}:
 \begin{itemize}
    \item The bracket indicates that we perform the Euclidean path integral over all fields in the gravitational theory including fluctuations of the metric.

    \item The boundary conditions are set as follows.
    As we will see in the beginning of Section~\ref{sec:zero-modes-in-Einstein-Maxwell-theory},  
    the asymptotics of the
    the near-horizon region of a near-extremal black with large but finite inverse temperature~$\beta$ is given by
    \be 
    \label{eq:asymptotic-metric}
    ds^2_{\rm asymp}= \ell_{AdS_2}^2 \underbrace{\left[d\rho^2 + \left(\rme^{2\rho}+\dots\right)d\tau^2 \right]}_{{\rm AdS}_2}+ \left(\ell_{S^2} + \delta \ell_{S^2} \rme^{\rho} + \dots \right)^2 \underbrace{\left[d\theta^2 + \sin^2 \theta d\phi^2\right]}_{S^2}\,,
    \ee 
    with 
    \be
   \rho>0,\,\qquad \tau \sim \tau+2\pi\,, \qquad \phi \sim \phi+2\pi\,, \qquad \theta\in [0, \pi]\,.
    \ee
    Here, the boundary is located at $\rho = \rho_c$ with $\rho_c \gg 1$, 
    and the allowed metric fluctuations in the functional integral are the subleading terms in an $\rme^{-\rho_c}$ expansion, shown by 
    the~$\dots$ in the equation above.  This implies that the proper length of the AdS$_2$ boundary  
    and the size $\ell_{S^2}+ \delta \ell_{S_2} \rme^{\rho_c}$ of the $S^2$ at the edge of the near-horizon region are fixed in all the metrics that we integrate over.

    The proper length of the AdS$_2$ boundary is given by $L = 2
    \pi\ell_{AdS_2}e^{\rho_c} = 
    \beta \ell_{AdS_2}/\epsilon$ where~$\beta$ can be identified as the inverse temperature of the black hole defined by the length of the Euclidean time circle at asymptotic flat space, and $\epsilon \, \sim\rme^{-\rho_c}$ is identified as a cut-off scale that sets the location of the AdS$_2$ boundary. As we shall see shortly, all observable quantities will be independent of the cut-off scale $\epsilon$.

    In previous treatments \cite{Sen:2008yk,Sen:2008vm} the size of $S^2$ is set to a constant, so that $\delta \ell_{S^2}\to 0$, and this leads to the presence of zero-modes. We will see that turning on a small  temperature makes~$\delta \ell_{S^2}$ non-zero, which  regulates these zero-modes.  
    Classically, the area of the $S^2$ slowly changes from its extremal value $4\pi \ell_{S^2}^2$ to its value $4\pi \ell_{S^2}^2+ 8\pi \ell_{S^2} \delta \ell_{S^2}\rme^{\rho_c} $ at the boundary.
    $\delta \ell_{S^2}$ can be conveniently parametrized by 
     rewriting $8\pi \ell_{S^2} \delta \ell_{S^2}\rme^{\rho_c} =  2G_N/\epsilon E_{SL(2)}$    where $E_{SL(2)}$ had been identified in the past as the temperature scale for the breakdown of thermodynamics close to extremality \cite{Preskill:1991tb}. 
    For a Reissner-N\"ordstrom black hole in flat space this energy scale is given by $E_{SL(2)} \sim  1/(Q^3)$, and consequently $\delta \ell_{S^2} = \pi Q^2 T$.

   A macroscopic black hole with an AdS$_2$ throat has 
   \be
  \frac{ \ell_{S^2}^2}{G_N}\; \gg\; \rme^{\rho_c}\;\gg \; 1\,,
   \ee
   and to consider its extremal limit in \eqref{qef} one takes 
    \be 
\frac{\ell_{AdS_2}\ell_{S^2}}{G_N}\frac{ \, \delta \ell_{S^2}\rme^{\rho_c}}{ L} \; \sim \; \frac{1}{\beta E_{SL(2)}} \; \to \; 0 \,,
    \ee    
    so that in the throat region the size of the $S^2$ is essentially unchanged. At finite  temperature it is also important to keep $\rme^{\rho_c} \ll (\ell_{S^2} T)^{-1}$ so that \eqref{eq:asymptotic-metric} give the correct asymptotics for the near-horizon region.

   \item In the second line in~\eqref{qef}, we have schematically included an integral over the metric, all the gauge fields (schematically labeled by $D A$) and all bosonic matter fields (schematically labeled by $D \chi$) or fermionic fields (schematically labeled by $D \Psi$ and including the spin-3/2 fields relevant in studying theories of supergravity) in the gravitational theory, whose integration measure we choose to be given by the ultra-local measure in the space of fields.

   \item Classically, the action $S_\text{grav, bulk}$ on AdS$_2$ is divergent and needs to be regulated. This is a standard procedure indicated by the upper-script ``reg" in the first line and realized  by including the appropriate boundary counter-terms in $S_\text{grav, bdy}$ in the second line. In addition to possible counter-terms,  $S_\text{grav, bdy}$
   includes the necessary boundary terms such that the gravitational theory has a well-defined variational principle. 
   For example in an Einstein-Maxwell theory this includes the Gibbons-Hawking-York term $S_\text{sugra, bdy} \supset \int d^4 x K$ where $K$ is the extrinsic curvature. Next, depending on the type of ensemble that we wish to study (grand-canonical or canonical) the associated boundary conditions for the gauge field at the AdS$_2\times S^2$ boundary could change. If we wish to study black holes in the grand-canonical ensemble, one imposes Dirichlet boundary conditions for the gauge field in the asymptotically flat region of a black holes geometry. This in turn implies that one has to fix an appropriate linear combination between the  gauge field and field-strength at the boundary of AdS$_2\times S^2$ in order to fix the chemical potential of the ensemble.  Here, we will instead study the canonical ensemble and fix the charge of gauge fields (by fixing the field-strength either in the asymptotically flat region or at the boundary of AdS$_2 \times S^2$ as we shall do below) instead of their chemical potential. To make the variational problem well defined for fixed charges requires (in any dimension) an electromagnetic boundary term $S_\text{sugra, bdy} \supset \int_{\partial(\text{AdS}_2\times S^2)} d^3 x \sqrt{h} A^\mu F_{\mu \nu} n^\nu$ \cite{Braden:1990hw, Hawking:1995ap}. After replacing the field strength in terms of the charges $q_I$ corresponding to each gauge field, this boundary term becomes precisely the Wilson loop insertion in the first line of \eqref{qef}.

\end{itemize}

In the classical limit for extremal black holes, the quantum entropy reproduces the Bekenstein-Hawking area term (or more generally the Wald entropy in the presence of higher derivative corrections) thus recovering the leading exponential in \eqref{eq:one-loop-result-structure-intro}. Performing the path integral described above beyond the classical approximation is an extremely complicated problem. Without attempting to reproduce the full black hole entropy, it is a reasonable question to ask what are the leading corrections to the classical answer. This can be obtained by expanding the gravitational action around the extremal black hole background (which we shall discuss in detail shortly but, for now, we denote by $\{g_{\mu\nu}^{\rm extremal}, A_\mu^{\rm extremal}, \dots\}$ where the dots are fields that are typically turned-off at extremality) and then by expanding the action to quadratic order in the fields 
\be 
I_{\text{grav}}[g_{\mu \nu}, A_{\mu}, \dots ]  &= I_{\text{grav, bulk}}[g_{\mu \nu}, A_{\mu}, \dots ]  + I_{\text{grav, bdy}}[g_{\mu \nu}, A_{\mu}, \dots ]\nn \\ &= I_{\text{grav}}[g_{\mu \nu}^{\rm extremal}, A_{\mu}^{\rm extremal},\, \dots ]+ \int d^4 x \sqrt{g} \left( \Phi K_b \Phi + \Psi K_f \Psi  \right) + \dots\,,
\ee 
where $\Phi$ and $\Psi$ are proxies for all bosonic and fermionic quantum fluctuations appearing in the quadratic expansion of the fields, $K_b$ and $K_f$ are the differential operators acting on these fluctuations and the ``\dots'' are higher-order fluctuations which are suppressed in the limit of weak gravitational coupling, $G_N \to 0$. 

After finding the bosonic/fermionic eigenfunctions of the differential operators $K_b$ and $K_f$, one finds two types of modes:
\begin{itemize}
    \item \textbf{Massive modes} that have non-zero eigenvalue under the action of $K_b$ and $K_f$. From these modes and their associated eigenvalues, one can construct a heat-kernel  of the operators $K_b^\text{bulk}$ and $K_f^\text{bulk}$, where the super-script indicates that we are considering modes that are not in the kernel of the differential operators. By extracting the constant part of the heat-kernel in a expansion for small values of the heat kernel parameter, \cite{Banerjee:2010qc,Banerjee:2011jp,Sen:2011ba,Sen:2012cj} managed to unambiguously recover the contribution of such modes to $c_{\rm log}$. 
    \item \textbf{Zero modes } of the kernel over $K_b$ and $K_f$ are either associated to large diffeomorphisms, large super-diffeomorphsism or large gauge transformations. 
    Each such transformation is normalizable, but is obtained by using a gauge parameter  (that is either a vector, spinor or scalar) that is non-normalizable~\cite{Camporesi:1995fb}.   
In the limit of zero-temperature in which the near-horizon region is precisely AdS$_2 \times S^2$ such modes are not only zero-modes at quadratic order, but exact zero modes of the gravitational action since their associated field strength vanishes. 
Thus, to compute their contribution to the gravitational path integral, we need to integrate over the entire moduli space of large gauge transformations. The dependence on the black hole area in the measure of this integral  was already understood in~\cite{Banerjee:2010qc,Banerjee:2011jp,Sen:2011ba,Sen:2012cj} and this accounts for the contribution of such zero-modes to $c_{\rm log}$. However, after stripping off the dependence on the area from the measure, the resulting integral over the moduli space of large gauge transformations is formally  divergent and therefore needs to be regulated. Since the moduli space of large gauge transformations is different in the case of extremal black holes in EM theory than in the case of extremal black holes in supergravity the regularization of this integral can yield drastically different corrections to \eqref{eq:one-loop-result-structure-intro}. 
\end{itemize}

For these reasons, in the remainder of this paper, we carefully go over the proper treatment of zero-modes appearing in near extremal charged black holes in four dimensions. We consider the case of Einstein-Maxwell and $\mathcal{N}=2,4,8$ supergravity.

\section{Treatment of zero-modes in Einstein-Maxwell}
\label{sec:zero-modes-in-Einstein-Maxwell-theory}

We begin with the analysis of Einstein-Maxwell gravity in four dimensions with action 
\beq
I \= \frac{1}{16 \pi G_N}\int d^4 x \sqrt{g} \left( R - F_{AB} F^{AB}\right) + \frac{1}{8\pi G_N} \oint \sqrt{h} K\,,
\eeq 
where $g$ is the 4d metric, upper-case indices label four-dimensional coordinates, $R$ is the scalar curvature and $F=dA$ is the gauge field strength. 
If we fix the charge of the gauge field there is an extra boundary term $\oint d\Sigma^A F_{AB} A^B$ necessary to make the variational problem well-defined~\cite{Braden:1990hw, Hawking:1995ap}. For simplicity we set $G_N=1$ from now on. 
The Euclidean metric of a charged Reissner-Nordstrom black hole solution is 
\beq
\label{eq:RNmetric}
ds^2 \= f d\tau^2 +\frac{dr^2}{f} + r^2(d\theta^2 + \sin^2 \theta d\phi^2),~~~f\=1-\frac{2M}{r}+\frac{Q^2}{r^2},~~~A\=-i\frac{Q}{r_+} \left( 1 - \frac{r_+}{r}\right)d\tau\,,
\eeq
where $M$ and $Q$ are the mass and charge of the black hole, while $r_+ = M + \sqrt{M^2-Q^2}$ is the location of the outer horizon. The temperature is given by $T=f'(r_+)/4\pi$ and arises from the identification $\tau \sim \tau+\beta$. The extremal black hole has zero temperature and corresponds to $M=Q$. To compute the quantum entropy one first zooms into the near horizon $AdS_2\times S^2$ region that develops close to extremality. 
In order to see this region we expand the metric above at small temperatures $TQ\ll 1$ as in~\cite{Sachdev:2019bjn}. 
We work at fixed charge $Q$ and the mass is then determined as a function of temperature by $M = Q + 4\pi^2 Q^3 T^2 + 16 \pi^3 Q^4 T^3 + \mathcal{O}(T^4)$. 
Upon introducing the new coordinates
\beq
r \to r_+(T) + 2\pi Q T (\cosh \rho-1),~~\qquad~\tau \to \frac{1}{2\pi T} \tau\,,
\eeq
plugging this replacement in the metric \eqref{eq:RNmetric}, expanding to leading order in $T$, and taking $r - r_+ \ll r_+$ (i.e.~$\rho\ll 1/T$), we obtain the $AdS_2\times S^2$ metric 
\be \label{eq:metricadss}
ds^2 \= Q^2 \left(d\rho^2 \,+\, \sinh^2 \rho\, d\tau^2  \right)+ Q^2 (d\theta^2 \, +\, \sin^2 \theta d\phi^2),~~~~A \=  i Q (\cosh \rho -1)d\tau\,.
\ee
From now on, $\mu,\nu,\ldots$ will label $AdS_2$ coordinates while $\alpha,\beta,\ldots$ the $S^2$ coordinates. In these coordinates the horizon is located at $\rho =0$. We see below we need to go to next order in the small temperature expansion, 
but we begin the discussion following \cite{Sen:2012cj} and considering quantum fluctuations around \eqref{eq:metricadss}. To compute the one-loop determinant one needs to integrate over fluctuations of the metric and gauge fields, 
written as~$g \to g + h$ and $A\to A+ \mathcal{A}$. The next step is to expand the Einstein-Maxwell action to quadratic order $I= I[g,A] + I_{\rm quad}[h,\mathcal{A}]+\ldots$, including proper gauge fixing terms and ghosts contributions \cite{Sen:2012cj}. 

\subsection{Massive modes}
\label{sec:massive-modes}

The quadratic approximation of the Einstein-Maxwell action involves differential operators acting on the metric and potential fluctuations. One can find a complete set of modes to expand these fluctuations such that the differential operators are diagonalized. These modes can be separated into `continuous' modes and `discrete' modes. The continuous modes are constructed in terms of the eigenfunctions of the scalar Laplacian, and are given by 
\beq
u_{\lambda,p;\ell,m}(x) \= f_{\lambda,p}(\rho,\tau)\, Y^\ell_m (\theta,\phi)\,.
\eeq
The first factor on the right-hand side is the eigenfunction of the Laplacian in AdS$_2$ with eigenvalue $\left(\frac{1}4 +\lambda^2\right)/Q^2$, with $\lambda \in \mathbb R^+$:
\be 
f_{\lambda, p}(\rho, \tau)&= \frac{1}{\sqrt{2\pi Q^2}} \frac{1}{2^{|p|}\Gamma(|p|+1)} \left|\frac{\Gamma\left(i \lambda + \frac{1}2 + |p|\right)}{\Gamma(i \lambda)}\right|\rme^{i p \tau} \sinh^{|p|} \rho \nn \\ &\times \,F\left(i \lambda + \frac{1}2 + |p|, -i \lambda +\frac{1}2 + |p|; |p|+1;-\sinh^2 \frac{\rho}2\right)\,.
\ee
 The second label $p\in \mathbb{Z}$ is quantized since we are working in coordinates where $\tau \sim \tau+2\pi$ and $p$ is the momentum in the angular direction $\tau$. The second term is the spherical harmonic $Y^\ell_m(\theta,\phi)$ on $S^2$ with eigenvalue $\ell(\ell+1)/Q^2$. Thus, these modes obey
\beq
-\Box \, u_{\lambda,p;\ell,m} \= \kappa_{\lambda,\ell} \, u_{\lambda,p;\ell,m},~~~\kappa_{\lambda,\ell} \= \frac{1}{Q^2}\left( \frac{1}{4}+\lambda^2 + \ell(\ell+1)\right).
\eeq
Using these functions we can construct a continuous set of modes of the Laplacian acting on metric or potential fluctuations. For example, as a normalized basis of vector fields on AdS$_2$ one can take $\partial_A u_{ \lambda,p;\ell,m}$ or $\varepsilon_{AB} \partial^B u_{\lambda,p;\ell,m}$, and similarly for $S^2$. It is also possible to write metric fluctuations as $g_{AB} u_{\lambda,p;\ell,m}$ or in terms of vector modes appearing as diffeomorphisms. 

A common feature of the continuous modes described in the previous paragraph is that they are not zero-modes of the Laplacian and, consequently, nor of the full action. 
This can be traced back to the fact that the scalar Laplacian eigenvalues are bounded by $\kappa_{\lambda,\ell}\geq (2Q)^{-2}$ due to the background curvature. Therefore, even  massless
scalar fields in the original theory have a non-zero Laplacian. The vector eigenvalues of such continuous modes are bounded below by $\frac{3}2 Q^{-2}$ and the metric modes by $\frac{9}2 Q^{-2}$. In the quadratic approximation of the Einstein-Maxwell action these different modes can mix with each other, but the resulting eigenvalues are found to still be bounded below by a non-zero constant \cite{Sen:2012cj}. The integration over all these modes produce temperature independent terms that are logarithmic in the area of the black hole. 

Besides the continuous modes, there are also discrete modes. These appear only in vectors and metric tensor. In the example of vectors, we have constructed the above modes made out of derivatives of normalizable scalar eigenfunctions of the Laplacian. 
The discrete vector modes are constructed from scalar functions which are not normalizable on~AdS$_2$, but produce a vector profile which is normalizable~\cite{Camporesi:1995fb}.
A similar statement can be made regarding tensor modes. Most of these discrete modes are massive as well, although a small number of them are exact zero-modes of the extremal metric, and the rest of this section is devoted to a careful analysis of the zero-modes.  

\subsection{Tensor zero-modes and the Schwarzian}
\label{sec:tensor-zero-modes-and-Schwarzian}

Discrete modes for metric fluctuations are diffeomorphisms along AdS$_2$ which are normalizable as metric fluctuations but the diffeomorphism itself is not, combined with a spin $\ell$ spherical harmonic in $S^2$. Out of these, it was found in \cite{Sen:2012cj} that the discrete mode with $\ell=0$ are exact zero-modes of the extremal metric and given by
\beq\label{eq:metrfldi}
h^{\varepsilon}_{AB}dx^A dx^B \= \sum_{n\in \mathbb{Z}, |n|\geq 2}2n(n^2-1) i  \varepsilon_n\rme^{i n \tau}\tanh^{|n|}\frac{\rho}{2}  \left(\frac{d\rho^2}{\sinh^2\rho} + \frac{2 i d\rho d\tau}{\sinh\rho} -  d\tau^2\right),
\eeq
for $n \in \mathbb{Z}$, such that $|n|\geq 2$. Since they do not mix with the rest of the fluctuations we can treat their contribution to the one-loop determinant separately, see \cite{Sen:2012cj}. These modes are precisely the Schwarzian sector arising from a reduction to two dimensional JT gravity, as explained for example in \cite{Iliesiu:2020qvm} using a different approach. First, as explained above, this metric fluctuation can be written as a pure diffeomorphism $x^A\to x^A+ \zeta^
A$ with
\beq
\zeta \=  \sum_{|n|\geq 2} \varepsilon_n\rme^{i n \tau} \tanh^{|n|} \frac{\rho}{2} \left(\frac{ i n (|n|+\cosh \rho)}{\sinh \rho} \partial_\rho -\frac{|n|(|n|+\cosh\rho)+\sinh^2\rho}{\sinh^2\rho} \partial_\tau\right).
\eeq
These diffeomorphisms act at infinity which can be seen by taking the $\rho \gg 1$ limit 
\bea
\zeta &\approx&  \sum_{|n|\geq 2} \varepsilon_n\rme^{i n \tau} \left(i n  \partial_\rho - \partial_\tau\right),\nonumber\\
&\approx&  \varepsilon'(\tau) \partial_\rho - \varepsilon(\tau) \partial_\tau,~~~~\varepsilon(\tau) \equiv  \sum_{|n|\geq 2} \varepsilon_n\rme^{i n \tau}.\label{eqn:reparamschwarzian}
\ea

In order to identify this function~$\varepsilon(\tau)$ with the Schwarzian mode we take the approach 
of~\cite{Maldacena:2016upp} and imagine considering rigid AdS$_2$ with a fixed length boundary. Then, the gravitational degree of freedom is encoded in the boundary curve which can be parametrized as $(\rho,\tau) \to (\hat{\rho},\hat{\tau})$ with  $\hat{\tau}=f(\tau)$ and $\hat{\rho}=\rho(\tau)$. 
To impose the Dirichlet boundary condition on the metric we set $[\rho'(\tau)]^2 + \sinh^2 \rho(\tau) [f'(\tau)]^2 = \sinh^2\rho_c$, where we denote the unrenormalized boundary length by $L=2\pi \ell_{{\rm AdS}_2} \sinh \rho_c$. For small fluctuations around a rigid boundary we take $f(\tau) = \tau + \varepsilon(\tau)$ with $\varepsilon(\tau)$ small. 
The fixed metric constraint gives then $\rho (\tau) = \rho_c - \varepsilon'(\tau) + \mathcal{O}(\varepsilon^2)$. This calculation was done in a coordinate system in which the metric is rigid AdS$_2$ but the boundary curved. We can change coordinates to a system more natural for the calculation outlined above with a rigid boundary but with fluctuations in the metric. In order to do this, it is required to perform near the boundary a diffeomorphism $\rho \to \rho + \varepsilon'(\tau)$ and $\tau \to \tau - \varepsilon(\tau)$ which 
agrees with the expression~\eqref{eqn:reparamschwarzian}. 

The upshot is that the large diffeomorphisms of $AdS_2\times S^2$ analyzed in the previous paragraphs can be parametrized by an element of ${\rm Diff}(S^1)/SL(2,\mathbb{R})$. The circle diffeomorphism is encoded in the function $\varepsilon(\tau)$. The modding by $SL(2,\mathbb{R})$ appears from the fact that diffeomorphisms with Fourier modes $n=-1,0,1$ produce a vanishing metric perturbation. The reason for this can be traced back to the isometries of AdS$_2$.

\subsection{Vector zero-modes and black hole rotations}
\label{sec:vector-zero-modes-and-rotations}

In addition to the tensor modes, there are additional vector discrete zero-modes in the metric. These can be interpreted as arising due to the rotation of the black hole, and the fact that they are zero modes is traced back to the isometries of $S^2$. Concretely, the metric deformation corresponding to them is given by
\be 
\label{eq:discrete-AdS2-vector-mode}
h_{\mu \alpha} &\= \sum_{|n|\geq1} \sum_{m=-1,0,1} v_{n, m}~ \epsilon_{\alpha \beta} \partial^\beta  Y_{\ell=1, m}(\theta, \phi)~\partial_\mu\left(\Phi_n(\rho, \tau) \right)\,, \nn \\ 
\Phi_n(\rho, \tau) &\= \frac{1}{\sqrt{2\pi |n|}} \left[\frac{\sinh \rho}{1+\cosh \rho} \right]^{|n|}\rme^{i n \tau}\,,
 \ee
 where $\epsilon_{\alpha \beta}$ is the Levi-Civita tensor in $S^2$ normalized such that $\epsilon_{\theta \phi} = Q^2 \sin \theta$. This can be shown to be a zero mode when $\ell =1$ and for any value of $n$. The sum over the coefficients $v_{n,m}$ is constrained by demanding the metric perturbation is real, just like with the $\varepsilon_{n}$ zero-modes. The relation between these zero-modes and the isometries of $S^2$ is clear since $\epsilon^{\alpha \beta} \partial_\beta Y_{\ell=1,m}$ are the three Killing vectors of the sphere \cite{Michelson:1999kn}. The rest of the right hand side can be interpreted as a pure gauge fluctuation of the $SU(2)$ gauge field arising in $AdS_2$ after reducing on the sphere. They can be interpreted as small fluctuations in the angular momentum of the black hole. One can see that these modes are orthogonal to all other zero and non-zero modes written down in~\cite{Sen:2012cj}.
 
We have seen that the tensor zero-modes $\varepsilon_n$ naturally parametrize the tangent space to ${\rm Diff}(S^1)/{\rm SL}(2,\mathbb{R})$. The functions $v_{n,m}$ have a similar interpretation. For a fixed $n$ the three components are associated to the three generators of $SU(2)$. Then the $v_{n,m}$ parametrize the tangent space to ${\rm Loop}(SU(2))/SU(2)$. The modding by a global $SU(2)$ arises from the fact that the fluctuation vanishes for $v_{0,m}$ and these three parameters generate the global symmetry. Just like we identified the tensor zero-modes with the Schwarzian theory emerging from JT gravity in the dimensional reduction of the original theory, we can similarly identify these modes as coming from a two dimensional $SU(2)$ Yang-Mills action in AdS$_2$ \cite{Iliesiu:2020qvm}. We would like to stress that the fact that these modes are massless depend on the boundary conditions and, in particular, we computed the action in the grand-canonical ensemble of fixed angular velocity. We do not expect them to be zero-modes if we fix the angular momentum instead.

\subsection{Vector zero-modes and gauge transformations}
\label{sec:vector-zero-modes-and-gauge-transf}

The last set of zero-modes come from the $U(1)$ gauge field. We can write fluctuations as pure gauge transformations that are not normalizable in $AdS_2$. More concretely it is given by 
\beq
\mathcal{A} \= d \lambda ,~~~~\lambda \= \sum_{|n|\geq 1} a_n~ \frac{1}{\sqrt{2\pi |n|}} \left[\frac{\sinh \rho}{1+\cosh \rho} \right]^{|n|}\rme^{i n \tau},
\eeq
where $a_n$ are constrained such that the fluctuation is real in Euclidean signature. It is clear in the expression above that these zero-modes come from an $\ell=0$ vector discrete mode. 
These modes now parametrize fluctuations around ${\rm Loop}(U(1))/U(1)$ with the global identification associated to the $n=0$ mode that is not being included. This is the same set of zero-modes that would arise from a $U(1)$ gauge field in AdS$_2$ in the grand-canonical ensemble.

For a reader familiar with \cite{Sen:2012cj} note that even though there is kinetic mixing between the discrete vector mode in the gauge potential and the metric, this mixing vanishes for the zero-modes corresponding to $\ell=0$. Therefore all zero-modes are orthogonal to all other zero-modes and to the non-zero modes written down in~\cite{Sen:2012cj}.

\subsection{Measure over the zero modes}
\label{sec:zero-modes-measure}

Next we need to determine the measure of integration over the modes generated by the diffeomorphisms $\zeta$. In order to do this we separate the calculation in two steps. The first step consists in identifying the dependence of the measure with the horizon area $Q$, which can be determined by scaling arguments. The second step consists in setting $Q=1$ and determining the measure over these large diffeomorphisms, since then we can quote results in the literature  that worked in those conventions.

The first step, i.e.~determining the overall $a$ dependence, was performed for example in~\cite{Sen:2012cj}. To determine the measure over these large diffeomorphisms we start with the measure over an arbitrary four dimensional metric fluctuation. The standard choice is to take the ultra-local measure defined implicitly through the equation 
\beq
\label{eq:ultra-local-measure}
\int Dh_{AB} ~e^{-\int d^4x\sqrt{g} g^{AB} g^{CD} h_{AB} h_{CD} } \= 1\,.
\eeq
If we insert the metric \eqref{eq:metricadss} into \eqref{eq:ultra-local-measure}, then its easy to see that the factors of~$Q$ cancel. Therefore the measure over~$h_{AB}$ has no explicit dependence on~$Q$. 
In order to relate this to the measure over the Schwarzian modes $\zeta \sim \varepsilon(\tau)\partial_\tau$ we need to incorporate the Jacobian appearing from integrating over $h_{AB}\sim D_{(A}\zeta_{B)}$ to integrating over $\zeta^A$. 
It is easy to see this adds an extra factor of $Q^2$ per zero-mode from lowering the index of $\zeta$. After fixing the dependence on~$Q$, we can use the result of~\cite{Moitra:2021uiv} which
computed the measure over $\zeta$ 
for~$Q=1$. Putting these two results together gives a measure $Dh_{AB} \to (D h_{AB})_{\rm non-zero}~\prod_{n\geq2} Q^4 ~ d\varepsilon_n d\varepsilon_{-n}|n|(n^2-1) $, up to numerical prefactor. Now we come to the main point of this discussion. While 
the~$Q$-dependence of the one-loop correction is correct both for non-zero and zero-modes, it is important to remember we need to perform the integral over~$\varepsilon(\tau)$ in the end. 

A similar analysis can be made to extract the $Q$ dependence of the vector zero-modes. At the end of the day the total one-loop determinant is given by
\beq
Z_\text{1-loop} \= Q^{c_{\rm log}} ~\int D \varepsilon~Dv~Da\,,
\eeq
where $v$ are the modes coming from an $SU(2)$ gauge field in AdS$_2$ and $a$ from the s-wave sector of $U(1)$ fluctuations. The parameter $c_{\rm log}$ is the coefficient of the logarithmic corrections to the entropy computed in \cite{Sen:2012cj} and the remaining integral over $\varepsilon,\,v$, and $a$ is independent of $Q$. For the example of Einstein-Maxwell coupled to matter this number is 
\beq
c_{\rm log}\=-\frac{1}{180}(964+n_S+62 n_V + 11 n_F)\,,
\eeq
where $n_S$ is the number of massless scalars, $n_F$ massless Dirac fermions and $n_V$ additional Maxwell fields besides the one giving the charge to the black hole. The final integral over the space of bosonic zero-modes is formally divergent since the space ${\rm Diff}(S^1)/SL(2,\mathbb{R})$ and ${\rm Loop}(G)/G$ for $G=U(1)$ and $SU(2)$ are all non-compact. This point is usually omitted in discussion of the quantum entropy function but is crucial to obtain the correct one-loop determinant when summing over orbifolds. In the next section we will explain a physical way to regulate this.

\subsection{Regularization of the zero-modes}
\label{sec:regularization-zero-modes}

Fluctuations around the $AdS_2 \times S^2$ metric given in \eqref{eq:metricadss} leads to divergences since it includes a non-compact space of zero-modes. A first prescription could be to simply remove these zero mode integrals by hand. This cannot be the correct answer since we expect the near-extremal entropy to grow linearly with temperature \cite{Preskill:1991tb} and this behavior comes precisely from regulating these zero modes \cite{Maldacena:2016upp}. 

Instead, the correct way of regularizing these zero modes is to go back to the full metric \eqref{eq:RNmetric} and expand to next order in $QT$. This modification of the background metric gives a temperature dependent mass 
to the zero-modes and at the same time incorporates near-extremal behavior of the entropy, solving both problems. To linear order in temperature the background metric is
\bea 
\delta g_{AB}dx^A dx^B &=& \pi Q^3 T (2+\cosh \rho)\tanh^2 \frac{\rho}{2}\left(d\rho^2 -  \sinh^2\rho d\tau^2  \right)+ 4\pi Q^3 T \cosh \rho (d\theta^2 + \sin^2 \theta d\phi^2),\nonumber\\
\delta A &=&-2i \pi Q^2 T \sinh^2 \rho d\tau.\label{eq:metricadsslinT}
\ea
This geometry should be cut-off at $\rho_c$ such that  $1\ll \cosh \rho_c\ll (QT)^{-1}$ otherwise the expansion of the metric is not accurate. It should also be taken into account when computing the action that there will be an order $T$ correction to the relation $L=2\pi Q \sinh\rho_c + \text{O}(T)$. One can check that the metric $g + \delta g$ and the potential $A+\delta A$ solve the Einstein-Maxwell equations to order $T^2$, and therefore is a good background to expand around. Of course this is immediate since we are expanding \eqref{eq:RNmetric} which is an exact solution for arbitrary temperatures. We should distinguish $\delta g$ (a change of the background) from $h$ (a quantum fluctuation of the metric around the background) and, similarly, $\delta A$ from $\mathcal{A}$. 

The first check we can make that this a physical prescription is to look at the evaluation of the classical action on the new saddle $g+\delta g$ and $A+\delta A$. The result gives a correction to the classical entropy given by
\beq
S \= \pi Q^2 + 4\pi^2 Q^3 T + \ldots,
\eeq
where the dots denote terms that are subleading either in the large-$Q$ expansion or the small-$T$ expansion.
This is precisely the expansion of the Bekenstein-Hawking area term at small temperatures. It also defines the scale $T \sim 1/Q^3$ at which there is a breakdown of the statistical description of the black hole \cite{Preskill:1991tb}. The modern interpretation is that this is the scale below which the zero modes become strongly coupled, as we will see below.

Now we can look at the $4d$ Einstein-Maxwell action for modes $h$ expanded around $g\to g+\delta g + h$. When computing the action there could in principle be terms proportional to $\int \delta g ~h$. These terms all vanish due to the fact that $g+\delta g$ solves the equations of motion to order $T$. 
The leading terms we will be interested in are quadratic around the perturbed solution (or cubic in terms of $AdS_2 \times S^2$) and have the form~$\int \delta g  h h$. 
Since $\delta g \sim T$ any mode that is a non-zero mode at zero temperature will just get a small correction to its mass giving a negligible contribution to the temperature dependence of the one-loop determinant. The situation is different for the discrete zero-modes mentioned above since to leading order their action is zero. For simplicity we consider now we are in the canonical ensemble and focus our discussion here on the tensor modes. We will discuss the role of vector modes, both gauge and rotational, in Section~\ref{sec:canonical-grand-canonical-ensemble}.   For tensor modes, the   action is now given by 
\beq\label{eqn:1loopwithschw}
Z_\text{1-loop} \= Q^{c_{\rm log}} ~\int D \varepsilon ~\exp{\left[-\frac{ T}{E_{SL(2)}} \sum_{n\geq 2} (2\pi)^2 (n^4-n^2)|\varepsilon_n|^2\right] }\,.
\eeq
where the energy scale $ E_{SL(2)}$ is found to be
\be E_{SL(2)} = 1/Q^3\,,
\ee
and can be thought of as the energy scale at which the one-loop determinants associated to~$\varepsilon_n$ become important. 
The most important aspect of the new exponential term is the fact that it is non-zero and of order~$Q^3 T$. It is also important to note the $n$-dependence, which completes the connection with JT gravity. This is precisely the Schwarzian action to quadratic order since 
\be 
\label{eq:Schwarzian-action}
\int d\tau ~{\rm Sch}\left(\tan{ \frac{\tau+\varepsilon(\tau)}{2}},\tau \right) \= \pi  -\frac{1}{2}\int d\tau (\varepsilon''(\tau)^2 - \varepsilon'(\tau)^2) + \mathcal{O}(\varepsilon^3)\,.
\ee
Going to Fourier space for the quadratic order term reproduces the action  \eqref{eqn:1loopwithschw} obtained from the 4d fluctuation in the background metric. In fact, the action~\eqref{eq:Schwarzian-action} can be 
reproduced to all orders in~$\varepsilon$ as can be seen by performing the dimensional reduction along $S^2$ in the near-horizon AdS$_2$ region~\cite{Iliesiu:2020qvm}. 
This is also guaranteed by requiring that the resulting action is invariant under $SL(2, \mR)$ transformations of $\tan\, \frac{\tau+\varepsilon(\tau)}2 $ which correspond to isometries that change the location of the boundary, but leave all physical observables invariant.

\subsection{The one-loop determinant around the near-extremal background}

Up to a numerical prefactor we can compute the integral using zeta function regularization, and the result is 
\beq
\label{eqn:finalentropy}
Z_\text{1-loop} \= Q^{c_{\rm log}}~ (Q^3 T)^{3/2},~~~~ S \= \pi Q^2+ 4\pi^2 Q^3 T + c_{\rm log} \log Q + \frac{3}{2} \log Q^3 T \,.
\eeq
The first two terms in the entropy are classical, the third term comes from the calculation of the quantum entropy function ignoring a proper treatment of zero-modes and the fourth term corrects this. We see that the effect of regulating the zero-modes introduces new $\log{Q}$ terms beyond the ones computed for example in \cite{Sen:2012cj}, besides the new $\log{T}$ term. 
For instance, the ratio of partition functions for black holes with the same small temperature but different charges is different than the one predicted by~\cite{Banerjee:2010qc,Banerjee:2011jp,Sen:2011ba, Sen:2012cj} for extremal black holes.

We see that non-supersymmetric extremal black holes do not exist since $T=0$ is a singular limit for the entropy and a vanishing limit for the partition function \cite{Iliesiu:2020qvm}. Another issue is that the new $\log Q$ terms seem to be in tension with the fact that one can compare the macroscopic calculation described in this section with the microscopic calculation from string theory constructions. 
Nevertheless its important to note that all these checks were performed for supersymmetric black holes. In that case,  as we will see next, 
both issues are resolved: all the $\log{Q^3T}$ terms cancel. This keeps  the $\log{Q}$ terms computed by the quantum entropy function unmodified and also make the space of extremal states well defined. 
For non-supersymmetric black holes this analysis predicts new $\log{Q}$ terms although there is nothing to compare it with.

\subsection{Working in the canonical or grand-canonical ensembles}
\label{sec:canonical-grand-canonical-ensemble}

So far we have discussed how the tensor-zero modes seen at zero temperature in Section~\ref{sec:tensor-zero-modes-and-Schwarzian} gain a weight when the black hole is raised to finite temperature, consequently leading to the $\log \,T$ corrections seen above. We have implicitly used the fact that the path integral over the other two sets of zero modes---the large gauge transformations (discussed in Section~\ref{sec:vector-zero-modes-and-gauge-transf}) as well as the vector zero-modes of the metric (discussed in Section~\ref{sec:vector-zero-modes-and-rotations})---give no other corrections in the canonical ensemble. Below we shall explain why that is the case and contrast this with the result in the grand canonical ensemble. In order to discuss the difference between the two ensembles, we need to discuss their associated boundary conditions in the gravitational path integral. As already mentioned in Section~\ref{sec:quantum-entropy-review}, in the canonical ensemble, we fix the field strengths, consequently fixing the gauge charges, and fix the components of the metric that fixes the angular momenta. In the grand-canonical ensemble, we fix the holonomy of the $U(1)$ gauge field and, consequently, the chemical potentials of the associated charges, and fix the components of the metric necessary to fix the angular velocity of the black hole. For both boundary conditions, we will start by fixing the boundary conditions at the asymptotically flat boundary of spacetime and derive the boundary conditions that we consequently need to impose at the AdS$_2 \times S^2$ boundary by classically solving the equations of motion for the gauge field and for the rotational degrees of freedom in the region in between the asymptotic and near-horizon boundaries. We shall call this the intermediate region. 

Since the analysis for both kinds of vector zero-modes is fairly similar, we shall focus on the case of the rotational zero-mode which will play a more important role when discussing theories of supergravity. To emphasize even more the similarity between these two modes, we note that fluctuations of the rotational vector zero-modes in \eqref{eq:discrete-AdS2-vector-mode} are equivalent to the fluctuations of the $SU(2)$ gauge field that arises from the isometry group of $S^2$ when dimensionally reducing to two dimensions. This is because  \eqref{eq:discrete-AdS2-vector-mode} can be more generally rewritten as 
\be 
h_{\mu \alpha} \= \sum_{m=-1, 0,1}{\mathbf{B}_{\mu}^a} ~{\xi_{\alpha,\, m}}\,,
\ee
on the entire spacetime, where $\xi_{\alpha,\,m} = \epsilon_{\alpha \beta} \partial^\beta Y_{\ell=1,\,m}(\theta, \phi)$ are the components of the Killing vector $S^2$, and where $\mathbf{B}_{\mu}^m T_m$ can be identified as an $SU(2)$ gauge field with $T_a$ the generators of $SU(2)$. In the near-horizon region, \eqref{eq:RNmetric} implies that those fluctuations of the metric can be identified as $\mathbf{B}_{\mu}^m = \sum_{|n|\geq 1} v_{n,m} \partial_\mu\left(\Phi_n(\rho, \, \tau)\right)$. Thus, we see that in the near-horizon region the zero-mode rotational vector fluctuation can be identified as large $SU(2)$ gauge transformations that change the boundary value of $\mathbf{B}_{\mu}^m$. It is most convenient to determine the boundary conditions for the modes $v_{n, m}$ in terms of $\mathbf B$ at the edge of the AdS$_2\times S^2$, by solving the equations of motion between in the intermediate region for $\mathbf B$. In a gauge where $\mathbf B$ is fixed to only have a non-zero $T^3$ component and in the coordinate system \eqref{eq:RNmetric}, this is given by \cite{Heydeman:2020hhw,Iliesiu:2020qvm}:\footnote{Above, we choose $T_i = 
\sigma_i/2$ as the generators of $SU(2)$, where $\sigma_i$ are the Pauli matrices.}
\be \label{eq:Bclasssol}
\mathbf{B}\= \ii \, \sigma_3 \left(\mC_1+ \frac{\mC_2}{r^3}\right) d\tau\,, \qquad H_{r\tau} \=-\ii \, \frac{3 \sigma_3 \mC_2}{r^4}\,,
\ee
where $H$ is the field strength associated to $\mathbf{B}$, while $\mC_1$ and $\mC_2$ are up to this point undetermined constants. Studying the two different ensembles implies \cite{Heydeman:2020hhw,Iliesiu:2020qvm}:
\begin{itemize}
    \item \textbf{The canonical ensemble. } In such a case $H_{r\tau}$ is fixed as $r \to \infty$ which in turn implies that $\mC_2$ is fixed. This in turn fixes the $SU(2)$ charge associated to $\mathbf B$ which can be identified as the angular momentum. Since $\mC_2$ is fixed, the field strength is also fixed at the boundary of AdS$_2\times S^2$, i.e.~$\delta H_{r\tau}|_{\partial AdS_2\times S^2} = 0$. As discussed in Section~\ref{sec:quantum-entropy-review}, such a boundary condition requires the addition of a boundary term under which once again the modes $v_{n,m}$ gain a weight when the black hole is placed at $T\neq 0$.
    
    \item  \textbf{The grand-canonical ensemble. } In such a case we fix the gauge field at $r \to \infty $ to be $\mathbf{B}_\tau = \ii \Omega_E \sigma_3/2$, where $\Omega_E$ can be identified as the Euclidean angular velocity of the black hole. This fixes the holonomy at the asymptotic boundary to be  $h = \exp\left(\ii \frac{\beta \Omega_E \sigma_3}2 \right)$.  This implies that $\mC_1= \Omega_E/2$ is fixed, while $\mC_2$ is not.  At the boundary of   AdS$_2\times S^2$ we can then solve for $\mC_2$ finding that the relation between $\mathbf{B}$ and $H$ is fixed at that location, $\left(\mathbf{B}_\tau - \ii r \frac{H_{r\tau}}3\right)|_{\partial AdS_2\times S^2} = \ii \frac{\sigma_3 \Omega_E}2$. In other words, in the grand-canonical ensemble we should no longer fix the value of the gauge field at the boundary of the near-horizon region -- we instead have the mixed boundary conditions $\delta\left(\mathbf{B}_\tau - \ii r \frac{H_{r\tau}}3\right)|_{\partial AdS_2\times S^2} = 0$. Once again, such a boundary condition requires the addition of a boundary term and the modes $v_{n,m}$ gain a weight when the black hole is placed at $T\neq 0$.
\end{itemize}

Following Sen, we want to compute the full gravitational path integral within the near-horizon geometry. It is useful then to translate the boundary conditions derived from \eqref{eq:Bclasssol} to a statement at the boundary of the $AdS_2 \times S^2$ throat. 
Write $r=Q + \delta r$ with $\delta r \ll Q$. Propagating the classical solution outside the throat gives the asymptotic behavior at the boundary of $AdS_2$:
\be \label{eq:Bclasssol2}
\mathbf{B}\; \sim \; \ii \sigma_3 \left(\tilde{\mC}_1 - \frac{3 \mC_2}{Q^4} \delta r \right) d\tau\,, \qquad H_{r\tau} \; \sim \; -\ii\frac{3 \sigma_3 \mC_2}{Q^4}\,,~~~~~~\tilde{\mC}_1 \=\mC_1 + \frac{\mC_2}{Q^3}.
\ee
This expansion shows the following important fact: Even though one may naively identify $\tilde{\mC}_1$ as the chemical potential, this is not true in AdS$_2$. Fixing the chemical potential corresponds instead to fixing $\mC_1$. To implement this on the boundary of AdS$_2$ we need to impose mixed boundary conditions as described in the previous paragraph.\footnote{This point is also important when considering the orbifolds studied in the companion paper \cite{Iliesiu:companionPaperMicrostateCounting}. These non-perturbative configurations shift $\tilde{\mC}_1$, but do not violate the mixed boundary conditions we propose in grand canonical ensemble, leaving $\mC_1$ unchanged. We thank A. Sen for raising this question. }

As for the Schwarzian modes $\varepsilon_n$, we can expand the Einstein-Maxwell action in terms of $v_{n,m}$ when $T$ is kept finite. The quadratic expansion of the action agrees with that of the $1$d action \cite{ Heydeman:2020hhw,Iliesiu:2020qvm, Anninos:2017cnw, Iliesiu:2019lfc},    
\beq
\label{eq:N=4-Schwarzian-action}
I_{SU(2)} =- \frac{T}{E_{SU(2)}} \int_0^{2\pi} d\tau  \,\,\Tr(g^{-1} \partial_\tau g)^2\,,
\eeq
where $g$ is an $SU(2)$ valued field (with the $v_{n,m}$ arising from expanding $g$ around the classical solution in Lie-algebra valued Fourier modes) and $E_{SU(2)}$ is found to be \cite{Heydeman:2020hhw,Iliesiu:2020qvm},\footnote{Note that this agreement seems to be simply a coincidence in Einstein-Maxwell theories in flatspace. We will see that this agreement will however play an important role in the theory of supergravity where it is necessary in order for the effective theory of superdiffeomorphisms to itself be supersymmetric.  } 
\be 
E_{SU(2)} \= E_{SL(2)} \= \frac{1}{Q^3}\,,
\ee
and where \eqref{eq:N=4-Schwarzian-action} can be obtained by identifying the gauge field on the boundary of AdS$_2\times S^2$, which is not fixed in either of the two boundary conditions,  as $\mathbf B = g^{-1}dg$. 
Just as in the case of the Schwarzian, the effective action \eqref{eq:N=4-Schwarzian-action} is found to capture all higher-order corrections from $v_{n,m}$ to the Einstein-Maxwell action. Thus, the path-integral over $g$ for the action \eqref{eq:N=4-Schwarzian-action} can be identified with that over the large diffeomorphisms that generate rotation around $S^2$ when going around the boundary of AdS$_2$. 

In the grand-canonical ensemble, the boundary condition for $\mathbf B$ implies that $g(\tau+2\pi) = h \cdot g(\tau)$, with $h =  \exp\left(\ii \frac{\beta \Omega_E \sigma_3}2 \right)$. The partition function of this mode can be written in such a case as, 
\be 
\label{eq:SU2-mode-part-function}
Z^{SU(2)}(\beta,\,\Omega_E) \= 
\sum_{J \, \in \, \frac12 \mathbb Z^{\ge 0}}(2J +1) \chi_J\left(\frac{\beta \Omega_E}{4\pi}\right)\rme^{-\beta\frac{E_{SU(2)}}2 J(J+1)}\,,
\ee
where $J$ can be identified as the angular momentum of the black hole and where we have defined the $SU(2)$ character $\chi_J(\alpha) = \sin (2J+1)2\pi \alpha / \sin 2 \pi \alpha$.
To go to the canonical ensemble with~$J=0$ we simply need to integrate \eqref{eq:SU2-mode-part-function} over the group elements $h$, using the $SU(2)$ Haar measure. This implies that the path integral in the canonical ensemble with $J=0$ associated to the modes $v_{n,m}$ is simply given by,
\be 
Z^{SU(2)}(\beta, J=0) \= 1\,.
\ee
Thus, this mode does not give rise to any corrections that are logarithmic  in the temperature, as implicitly assumed 
when writing \eqref{eqn:finalentropy}. An identical analysis for the $U(1)$ gauge fields reveals that when setting all the charge fluctuations associated to large gauge transformations to vanish ($\tilde q=0$), the path integral over such modes also have $Z^{U(1)}(\beta, \tilde q=0) = 1$ which again verifies that it was correct to ignore the contribution of large gauge transformations in  \eqref{eqn:finalentropy}.

\subsection{Regime of validity of $\log T$ corrections}

We have computed the full one-loop correction to the black hole entropy including the boundary zero-modes. Interactions between non-zero modes are suppressed in powers of $Q^2$. 
Interactions between the large boundary zero-modes instead have a coupling which grows as an inverse power of temperature. In this section we comment on the regime of validity of the one-loop temperature dependence obtained in \eqref{eqn:finalentropy}. 

The zero-mode metric fluctuation is denoted by $h^{\varepsilon}_{AB}$, and is given in~\eqref{eq:metrfldi}. We start by considering self-interactions of the zero-modes. Turning on temperature gives them a mass obtained in equation \eqref{eqn:1loopwithschw}. To leading order, in a small temperature expansion, all terms in the zero-mode self-interactions are proportional to $Q^3 T$. After rescaling $ h^\varepsilon \to (Q^3 T)^{-1/2} \tilde{h}^\varepsilon$, the action has the schematic form
\beq\label{eq:expansionselfinteractioneps}
I \= Q^3 T \left( \int K_2(h^\varepsilon)^2 + \sum_{n>2} c_n \int K_n(h^\varepsilon)^n\right)
\; \to \;  \int K_2(\tilde{h}^\varepsilon)^2 + \sum_{n>2} \frac{1}{(Q^3 T)^{\frac{n-2}{2}}}c_n \int K_n(\tilde{h}^\varepsilon)^n.
\eeq
Here,~$K_n$ denote the appropriate kernels appearing in the action involving $n$ nearly zero-modes, whose explicit form will not be relevant other than the fact that they are temperature independent. This shows that the self-interactions become large as $Q^3 T\to0$.

The observations in the previous paragraph would naively imply that the result~\eqref{eqn:finalentropy} is only accurate for $Q^3 T \gg 1$ but, as argued in \cite{Iliesiu:2020qvm}, this conclusion is incorrect. The action involving leading-order self-interaction between the zero-modes is precisely the Schwarzian action, in terms of the function $f(\tau) = \tau + \varepsilon(\tau)$ 
(as discussed below equation \eqref{eq:Schwarzian-action}). This theory is 
solvable~\cite{Stanford:2017thb, Mertens:2017mtv,Lam:2018pvp} and known to be one-loop exact. 
For this reason, all contributions from nearly zero-mode self-interactions 
to~\eqref{eqn:finalentropy} vanish except the one-loop contribution, {i.e.}~the quadratic term in~\eqref{eq:expansionselfinteractioneps}, 
indicating that the result is valid for temperatures such that $Q^3 T \sim \mathcal{O}(1)$.

The only remaining terms to consider are the ones that mix non-zero modes with nearly zero-modes. This was addressed in \cite{Iliesiu:2020qvm} as well \footnote{See also \cite{Maldacena:2019cbz, Ghosh:2019rcj,Maxfield:2020ale} for similar analysis in other contexts.}. This interaction can be obtained in the following way: first, compute the non-zero mode contributions in AdS$_2$ (including all KK modes on S$^2$) with a wiggly boundary parametrized by the function $\varepsilon(\tau)$ as explained below \eqref{eqn:reparamschwarzian}. This gives the effective action for the Schwarzian mode after integrating out zero-modes. When the non-normalizable component of non-zero mode fluctuations near the AdS$_2$ boundary is set to zero (such that we are not turning on sources besides the ones in the classical black holes background) the matter partition function is independent of the Schwarzian mode, up to corrections subleading in the cut-off \footnote{An explicit example of this phenomenon is given in Appendix C of \cite{Yang:2018gdb}, where the first subleading correction in the cut-off is computed.}.

We considered all perturbative corrections to the entropy and showed they are subleading in the extremal limit besides the ones contained in \eqref{eqn:finalentropy}. 
This suggests that the degeneracy of
non-supersymmetric extremal black holes vanishes since 
\beq
Z \; \sim \; \rme^{\pi Q^2 + \text{O}(1/Q^2)} Q^{c_{\rm log}} (Q^3 T)^{3/2} \to 0, ~~~~~~~\text{as}~T\to 0.
\eeq
However, this is not entirely accurate: in this limit, as the partition function becomes arbitrarily small  non-perturbative corrections can compete with the contribution from the black hole saddle (and its perturbative corrections). 
One example is provided by a geometry where a spacetime wormhole is added inside the bulk of AdS$_2$, whose classical action is zero at large $Q$. 
For this reason we expect \eqref{eqn:finalentropy} to be valid in the range
\beq
Q^{\beta} \rme^{-\alpha Q^2 } \ll  T \ll 1/Q\,,
\eeq
where $\alpha$ and $\beta$ are order one numbers which could be computed with more knowledge of the possible non-perturbative corrections. In terms of the black hole density of states, \eqref{eqn:finalentropy} begins to fail when the energy above extremality is so small that the perturbative correction predicts an order one number of extremal black hole states. The upper bound comes from demanding the near-extremal approximation to be accurate but, since we are interested in the small temperature limit anyway, this is automatically satisfied.

\section{Treatment of zero-modes in supergravity}
\label{sec:zero-modes-in-SUGRA}

\subsection{Differences with the non-supersymmetric case}

We being by analyze half-BPS black holes in $\mathcal{N}=2$ supergravity. The bosonic part of the action is the same as before, and therefore the black hole solution is also the same. The only modification in the one-loop determinant is the presence of a contributions from the two Majorana spinors $\Psi^{1}_A$ and $\Psi^{2}_A$. To quadratic order in the fermions, their action is 
\beq
\frac{1}{16\pi } \int d^4x \left[-\frac{1}{2} \bar{\Psi}^i_{A} \Gamma^{ABC} D_B \Psi_C^i + \frac{1}{2} F^{AB} \epsilon_{ij} \Psi^i_A \Psi^j_B + \frac{1}{4} F_{CD} \bar{\Psi}^i_{A} \Gamma^{ABCD} \epsilon_{ij}\Psi^j_B\right],
\eeq
where we defined the anti-symmetric matrix $\epsilon_{12}=-\epsilon_{21}=1$. As explained in \cite{Sen:2012cj} some gauge fixing terms and ghosts are required. As analyzed in the same reference spin 3/2 fields can be expanded in continuous and discrete modes of AdS$_2$, both multiplied by a spinor spherical harmonic on the sphere. The normalizable discrete modes is constructed from derivatives of spin 1/2 fluctuations which themselves are not normalizable. Out of the discrete modes, there is a set of normalizable zero modes of the gravitino which we analyze next.

We are not going to write them down explicitly, but a careful analysis of the action done in \cite{Sen:2012cj} shows there are four set of real gravitino zero-modes labeled by an integer. These can be written as a local supersymmetry with parameter $\epsilon(x)$ such that $\delta \Psi^i_A = D_A \epsilon + \ldots$ and that is non-normalizable in $AdS_2$. The sphere components of the zero-modes come from a spin-1/2 spherical harmonic. For example there is a complex mode $\epsilon_+ \sim\rme^{ i \phi/2}\rme^{i k \tau} \epsilon_0(\theta,\rho)$ and $\epsilon_- \sim\rme^{-i \phi/2}\rme^{i k \tau} \epsilon_0(\theta,\rho)$, for $k\in \mathbb{Z}$ and $\epsilon_0$ is a four dimensional spinor that can be determined from \cite{Sen:2012cj}. We see then that the gravitino zero-modes transform as the fundamental of $SU(2)$ from the AdS$_2$ perspective. Moreover, modes with $k=\pm 1/2$ are not included in the one-loop calculation since they generate a vanishing gravitino fluctuation $\delta \Psi^i_A = 0$. These are precisely the eight Killing spinors of the extremal black hole. The bosonic isometries together with the fermionic ones make the superconformal group $PSU(1,1|2)$. The analysis of \cite{Heydeman:2020hhw} then shows that the metric zero-modes and gravitino zero-modes combine into two dimensional $\mathcal{N}=4$ JT gravity in AdS$_2$. The contribution from non-zero-modes and from the $Q$ dependence of zero-modes in the case of $\mathcal{N}=2$ supergravity is 
\beq
c^{\mathcal{N}=2}_{\rm log} \= (23 + n_H - n_V)/12\,,
\eeq
when coupled to $n_H$ hypermultiplets and $n_V$ vector multiplets that are massless in four dimensions. 

The case of quarter BPS black holes in $\mathcal{N}=4$ supergravity or one eighth BPS black holes in $\mathcal{N}=8$ can be analyzed in a similar fashion, see \cite{Banerjee:2011jp}. The spectrum of non-zero modes is certainly different than the $\mathcal{N}=2$ case since the field content in these theories is different. Nevertheless, the work to determine $c_{\rm log}$ was already done in those references. On the other hand, the spectrum of zero modes from the metric and gravitino is universal. In all case its found a set of Schwarzian modes, $SU(2)$ modes, and the same number of fermionic zero-modes from the gravitino transforming in the fundamental of $SU(2)$. This implies the zero-modes in all these theories can be regulated via $\mathcal{N}=4$ JT gravity, which we review next. We work in the micro-canonical ensemble for any $U(1)$ gauge field which removes the logarithmic in temperature contribution, allowing us to focus on the Schwarzian mode.

\subsection{Regularized zero-modes}
\label{sec:one-loop-determinant-from-superSchw}

We have identified an emergent set of ${\rm Diff}(S^1)/SL(2,\mathbb{R})$ in the four dimensional path integral computing the partition function of near extremal black hole in Einstein-Maxwell theory. In the case of $\mathcal{N}=2,4,8$ supergravity, these modes are paired up with $SU(2)$ modes and four fermionic modes, and together make the group
\beq
\text{Space of nearly zero-modes:}~~~~{\rm Diff}(S^{1|4})/PSU(1,1|2)\,.
\eeq
We now want to give the calculation of the temperature dependence of the zero-modes. We have verified that small temperature corrections gives a non-zero action to the bosonic modes reproducing the Schwarzian action. Due to supersymmetry the same should happen with all modes reproducing the $N=4$ Schwarzian theory studied in \cite{Heydeman:2020hhw}. We leave a more explicit derivation of this fact for future work.

The $N=4$ Schwarzian theory has a bosonic reparametrization mode $f(\tau)$ a time-dependent $SU(2)$ transformation $g(\tau)\in SU(2)$ and four fermions $\eta(\tau)$ organized in a complex fundamental representation of $SU(2)$. The action of the theory is 
\beq
I_{N=4} \=- C \int_0^{2\pi} d\tau \left[\Sch(f,\tau) + \Tr(g^{-1} \partial_\tau g)^2 + ({\rm fermions})\right],
\eeq
where $C$ is the coupling constant which we already computed to be $C=Q^3 T$. The calculation of the one-loop determinant for this theory was done in \cite{Heydeman:2020hhw}. This calculation is easier to do in the grand canonical where we fix the angular velocity instead of angular momentum, and therefore we are computing the contribution to ${\rm Tr} \left[\rme^{-\beta H + 4 \pi i \alpha J}\right]$, where $J$ is the angular momentum along the $S^2$ direction. To simplify notation, we use $\alpha$ to parameterize the angular velocity in the following way
\begin{equation}
\Omega_E \= \frac{4 \pi \alpha }{\beta}. 
\end{equation}
The answer for the nearly zero-modes path integral as a function of temperature and angular velocity (through $\alpha$) is given by
\be 
\label{eq:localization-part-function}
Z_{\text{Nearly Zero-Modes, GCE}}(\beta,\alpha)  &\= \sum_{n \in \mathbb Z} \det_{\text{Schw., one-loop}} \,\,\det_{SU(2), \text{ one-loop}}\,    \det_{\text{ferm., one-loop}}\rme^{-I_{\cN=4,\text{bosonic}}^{\,\text{on-shell}} } \nn \\ 
&\=\sum_{n \in \mathbb Z} \frac{1}{C}\frac{2\cot(\pi \a) (\a+n)}{\pi^3 (1- 4(n+\a)^2)^2}\rme^{2\pi^2 C \left(1 -4  (n+\a) ^2 \right)}.
\ee 
The integer $n$ labels a set of classical solutions related by shifting the angular velocity by $\Omega \to \Omega + 4\pi i n/\beta$. The fact that the one-loop determinant has an overall factor of $1/C$, where $C=Q^3 T$, can be understood by counting gauged zero-modes. For the $\mathcal{N}=4$ Schwarzian theory there are six bosonic zero-modes from both $SL(2,\mathbb{R})$ and $SU(2)$ while there are eight fermionic zero modes, giving a factor of $\sqrt{C}^{\#_B - \#_F}=1/C$. This result might seem counter intuitive since it suggests the partition function diverges in the zero temperature limit, while we expect to obtain a finite answer from the counting of extremal black holes. The way this works is that this divergence is removed by the sum over saddles obtained by integer shifts of the angular velocity. 

It is instructive to rewrite the result in the canonical ensemble of fixed angular momentum, as explained in \cite{Heydeman:2020hhw}. The spectrum can be expanded in $\mathcal{N}=4$ supermultiplets $(J)\oplus 2(J-\frac{1}{2}) \oplus (J-1)$ and is given by
\bea
\label{eq:N4-dos-super}
\hspace{-0.4cm}Z_{\text{Nearly Zero-Modes, GCE}}(\beta,\alpha) \= 1 + \int_0^\infty dEe^{-\beta E}\sum_{J\geq 1/2} \left(\chi_J(\alpha)+ 2 \chi_{J-\frac{1}{2}}(\alpha)+\chi_{J-1}(\alpha) \right) \rho_J(E) \,,
\ea
where we defined the $SU(2)$ character $\chi_J(\alpha) = \sin (2J+1)2\pi \alpha / \sin 2 \pi \alpha$. The density of states per supermultiplet labeled by the highest spin $J$ is given by 
\beq
\rho_J(E) \= \frac{J \sinh \Big(2 \pi \sqrt{2 Q^3(E-E_0(J))}\Big)}{4\pi^2 Q^3\rme^2}~\Theta(E-E_0(J))~~~E_0(J)\= \frac{J^2}{2Q^3}.
\eeq
In this expression the density of states for each supermultiplet starts contributing at a gap scale $E_0(J)$. Since the continuum in \eqref{eq:N4-dos-super} starts at $J=1/2$ we see the smallest gap is given by $E_{\rm gap} = 1/(8 Q^3)$. On the other hand, the first factor of one in the first line of \eqref{eq:N4-dos-super} corresponds to both zero energy (measured with respect to extremality) since its independent of temperature, and zero angular momentum since its independent of angular velocity. This is the type of term required to make the extremal degeneracy survive at zero temperature. This degeneracy, since the grounds state have $J=0$ is the same as the index. The contribution of the nearly zero-modes to the index can be computed as the grand canonical partition function with a specific angular velocity $\alpha=1/2$ since $(-1)^F=e^{2\pi i J}$. Then in the limit $T\to 0$ we can compute 
\bea
Z_{\text{Nearly Zero-Modes, GCE}}(\beta\to\infty,\alpha)&\=&1+\text{O}(\rme^{-\beta/Q^3 } ), \\
Z_{\rm Index} \= Z_{\text{Nearly Zero-Modes, GCE}}(\beta,\alpha=1/2)&\=&1.
\ea
The most straightforward way to check that the contribution to the index from the nearly zero-modes is temperature independent and non-zero is to verify in \eqref{eq:N4-dos-super} that the supermultiplet character vanishes with the corresponding angular velocity and only the first line survives. This verifies the general arguments given in \cite{Dabholkar:2010rm} that the index should match the degeneracy. Using this result we can obtain the zero temperature contribution to the one-loop determinant
\bea
Z_{\text{Nearly Zero-Modes}}(J) &\=& \int_0^{1} d\alpha ~\chi_J(\alpha)~Z_{\text{Nearly Zero-Modes, GCE}}(\beta\to\infty,\alpha),\\
&\=& \delta_{J,0}.
\ea
The final answer for the one-loop contribution to the quantum entropy function from all fields is then 
\bea
\label{eq:one-loop-sugra-zero-modes}
Z_{\rm 1-loop}(J) &\=&  Q^{c^{\mathcal{N}=2}_{\rm log}} \int D(\text{Nearly Zero-modes}),\\
&\=& Q^{c^{\mathcal{N}=2}_{\rm log}} ~\delta_{J,0}.
\ea
This puts the results of the quantum entropy function on firmer grounds. Even though we found no difference for the case of supersymmetric Reissner-N\"ordstrom black holes, the total one-loop determinant of other black holes can be completely different than expected. For example we saw that for non-supersymmetric black holes $Z_{\rm 1-loop} = 0$, signaling that quantum effects implies there are no extremal black holes. The same is true for black holes with isometries $OSp(1|2)$ near the horizon. An intermediate case is given by black holes with an emergent isometry $SU(1,1|1)$ near the horizon, relevant for gauged supergravity. In this case the extremal black hole space can be destroyed or not by quantum effects depending on parameters of the model, as emphasized in \cite{Boruch:2022tno}.

\subsection{Index versus degeneracy}

The one loop determinant \eqref{eq:one-loop-sugra-zero-modes} shows that supersymmetric black holes have a large degeneracy that is entirely made up of microstates with $J=0$. It is informative to 
check that this indeed agrees with the computation of the supersymmetric index of the black hole, i.e.~we would like to compute $\Tr \,(-1)^F \rme^{-\beta H}$ in the putative black hole Hilbert space.
This can be achieved by studying the grand-canonical ensemble for angular momenta (keeping the ensemble canonical for all gauge charges) with an angular velocity $\Omega_E = 2\pi/\beta$ (or, equivalently, $\alpha=1/2$), such that $\Tr\, \rme^{-\beta H + i \beta \Omega_E J } \to \Tr\, \rme^{-\beta H +2\pi i J} = \Tr (-1)^F\rme^{-\beta H}$. Considering such an angular velocity does not affect the boundary condition for the non-zero modes 
since fermionic fields remain anti-periodic around the contractible cycle in AdS$_2\times S^2$. As described in Section~\ref{sec:canonical-grand-canonical-ensemble}, what changes however is are the boundary conditions for the vector-zero mode of the metric describing rotations on~$S^2$. 
The partition function of large super-diffeomorphisms (which include such a rotational mode) can be computed for arbitrary $\alpha $. When $\alpha = 1/2$, only the $n=0$ and $n=-1$ saddles survive in~\eqref{eq:localization-part-function}~\cite{Iliesiu:2021are} and, after summing the contributions of the two saddles and their one-loop determinants, one consequently confirms that 
\be 
Z_{\text{Nearly Zero-Modes}}(T\to 0, \,J=0) 
&\=  Z_{\text{Nearly Zero-Modes}}\left(T\to 0, \,\alpha \= 1/2\right)\,, \nn \\ \qquad
\Rightarrow Z_{\rm 1-loop}(T\to 0, \,J=0) &\=  Z_{\rm 1-loop}\left(T\to 0, \,\alpha = 1/2\right) \= Q^{c_{\rm log}} \,, 
\ee
or, in other words, the index and degeneracies of the black holes are indistinguishable. In the companion paper \cite{Iliesiu:companionPaperMicrostateCounting}, we confirm that this equality holds not only as above (when studying the non-zero modes at a quadratic order  while integrating over the zero-modes exactly) but,  indeed, for the exact degeneracy and index at least in the case of~$1/8$-BPS black holes in~$\cN=8$ supergravity in flat space.

\section{Conclusions}

We have revisited the computation of quantum corrections around extremal black holes, placing a focus on the role of large diffeomorphisms in the $AdS_2 \times S^2$ throat. While generic modes generate corrections logarithmic in the black hole entropy $\log S_0$, these zero-modes generate corrections logarithmic in the temperature $\log T$. The Seeley-DeWitt expansion of the heat kernel allows to extend the evaluation of the $\log S_0$ corrections to other spaces besides $AdS_2 \times S^2$. It would be interesting to develop a similar expansion in a low temperature limit that computes $\log T$ corrections in general spaces\footnote{We thank A. Castro and V. Godet for discussions.}. The arguments of \cite{Iliesiu:2020qvm,Heydeman:2020hhw} would indicate that $\log T$ corrections are more universal and depend only on the amount of symmetry emerging near the horizon as we take the small temperature limit. We hope to address this in future work.

\subsection*{Acknowledgements} 
We thank A. Castro, M. Heydeman, V.~Godet, R.~Gupta, J.~Simon, and W. Zhao for valuable discussions. GJT is supported by the Institute for Advanced Study and the National Science Foundation under Grant No. PHY-1911298, and by the Dipal and Rupal Patel funds. LVI was supported by the Simons Collaboration on Ultra-Quantum Matter, a Simons Foundation Grant with No. 651440. 
This work is supported by the ERC Consolidator Grant N.~681908, ``Quantum black holes: A macroscopic 
window into the microstructure of gravity'', and by the STFC grant ST/P000258/1. This work was performed in part at the Aspen Center for Physics, which is supported by National Science Foundation grant PHY-1607611.

\appendix

\bibliographystyle{utphys2}
{\small \bibliography{Biblio}{}}

\end{document}